\documentclass[reprint,eqsecnum,floats,aps,amsmath,amssymb,nofootinbib,prd,onecolumn, showpacs]{revtex4-1}
\usepackage{graphicx,physics}
\usepackage{amsmath,amssymb,mathtools,mathrsfs}
\usepackage{hyperref}
\usepackage{graphicx}
\usepackage{subfigure}
\usepackage{arydshln}
\usepackage{xcolor}
\usepackage{braket}
\usepackage{tensor}
\usepackage{enumitem,array,textcomp}

\begin{document}
	
	\title{Exploring alternatives to the Hamiltonian calculation of the Ashtekar-Olmedo-Singh black hole solution}
	\author{Alejandro Garc\'ia-Quismondo}
	\email{alejandro.garcia@iem.cfmac.csic.es}
	\affiliation{Instituto de Estructura de la Materia, IEM-CSIC, Serrano 121, 28006 Madrid, Spain}
	\author{Guillermo  A. Mena Marug\'an}
	\email{mena@iem.cfmac.csic.es}
	\affiliation{Instituto de Estructura de la Materia, IEM-CSIC, Serrano 121, 28006 Madrid, Spain}
	\begin{abstract}
		In this article, we reexamine the derivation of the dynamical equations of the Ashtekar-Olmedo-Singh black hole model in order to determine whether it is possible to construct a Hamiltonian formalism where the parameters that regulate the introduction of quantum geometry effects are treated as true constants of motion. After arguing that these parameters should capture contributions from two distinct sectors of the phase space that had been considered independent in previous analyses in the literature, we proceed to obtain the corresponding equations of motion and analyze the consequences of this more general choice. We restrict our discussion exclusively to these dynamical issues. We also investigate whether the proposed procedure can be reconciled with the results of Ashtekar, Olmedo, and Singh, at least in some appropriate limit.  
	\end{abstract}
	
	\maketitle					

\section{Introduction}

Over two years ago, a new model to describe black hole spacetimes in effective loop quantum cosmology was put forward in Refs. \cite{AOS1,AOS2,AO} by Ashtekar, Olmedo, and Singh (AOS). The work of these authors is set apart from previous related investigations in the literature (see Refs. \cite{1,2,3,4,5,6,7,8,9,10,11,12,13,14,15,16,17,18,19,20,21,22,23,24,25,26,27,28,29,30,31,32,33,34,35,36}, among others) owing to a combination of features. On the one hand, the main focus is placed on black hole related aspects rather than issues central to anisotropic cosmologies. On the other hand, the resulting model is claimed to display neither a dependence on fiducial structures nor large quantum effects on low curvature regions. By virtue of the introduction of quantum geometry (QG) effects, which is implemented by means of two polymerization parameters,\footnote{The term polymerization refers to the name ``polymer quantization'', which is often employed for the quantization of symmetry reduced models with loop techniques. In a nonformal way, this denomination can be thought of as being related to the 1-dimensional nature of the quantum excitations of the gravitational field, which implies that this field lives on the edges of 1-dimensional graphs, leading to this polymer-like picture of spacetime geometry.} the classical singularity at the center of the black hole is replaced with a transition surface that joins a trapped region to its past and an anti-trapped one to its future, extending the Schwarzschild interior to encompass what is interpreted as a white hole horizon. The resulting metric, that we will call effective in the sense that it can be treated classically but incorporates QG modifications, is smooth and its curvature invariants admit upper bounds that do not depend on the mass of the black hole \cite{AOS1,AOS2,AO}. The model is completed with a description of the exterior region that can be joined smoothly to the interior region, both to its past and its future, resulting in an extension of the whole Kruskal spacetime. 

The authors of Refs. \cite{AOS1,AOS2,AO} emphasize that they adopt a mixed prescription for the implementation of the improved dynamics. Indeed, instead of choosing the relevant polymerization parameters as constants or as arbitrary phase space functions, they claim to fix them to be Dirac observables. However, they do not treat them as such in their Hamiltonian calculation: in practice, the polymerization parameters are regarded as \emph{constants} in that calculation and, once the dynamical equations have been derived and solved, the parameters are set equal to the value of certain functions of the ADM mass of the black hole, which is a Dirac observable itself. This fact was already noted by Bodendorfer, Mele, and M\"unch  in Ref. \cite{N}, where they showed that a genuine treatment of the polymerization parameters as constants of motion, which are constant only along dynamical trajectories (i.e. on shell) but not on the whole phase space, would produce an extra phase-space dependent factor in the Hamiltonian equations. The analysis carried out in Ref. \cite{N} exploits the structure of the Hamiltonian constraint of the system, which is composed by the difference of two Dirac observables (the on-shell value of each of which turns out to be the black hole mass), to divide the phase space into two independent subsectors, associated with the degrees of freedom along the radial and angular spatial directions. In each subsector, the dynamics is generated by one of these constants of motion, which can then be regarded as \emph{partial Hamiltonians}. Additionally, in Ref. \cite{N} these Dirac observables play the role of polymerization parameters, in the sense that each of the parameters is taken to be a function only of its associated partial Hamiltonian. On shell, this is equivalent to deal with parameters that are functions of the black hole mass, and at least in this sense one would recover the original proposal of Refs. \cite{AOS1,AOS2,AO}. 

Nonetheless, since the two partial Hamiltonians become equal by virtue of the vanishing of the constraint, there is no telling apart which of the two contributes to the on-shell expression of each of the polymerization parameters. Therefore, one may argue that each parameter should be taken as a function of \emph{both} partial Hamiltonians, something that breaks the decoupling of subsectors at the Hamiltonian and dynamical levels. In the following, we focus our discussion exclusively on examining whether there exists an alternative procedure to carry out the Hamiltonian calculation starting from this observation, leaving apart other issues related with the asymptotic behavior of the metric, its physical interpretation, or quantum covariance, that are beyond the scope of this work (for recent criticisms on the AOS viewpoint on these issues, see Refs. \cite{Bouhmadicrit,Bojocrit,Bojo1,Arruga}). The main purpose of our investigation is to explore the possibility that one can develop an alternative dynamical analysis based on the cross-dependence of the polymerization parameters on the two partial Hamiltonians of the system, and study whether this possibility can reconcile in some sense the derivation of the solution presented in Refs. \cite{AOS1,AOS2,AO} with a genuine consideration of the parameters as constants of motion.

The rest of this paper is structured as follows. In Sec. \ref{sec:eq} we explore the consequences of polymerization parameters that are functions of both partial Hamiltonians as regards the derivation of the equations of motion associated with the Hamiltonian constraint of Refs. \cite{AOS1,AOS2,AO}. In Sec. \ref{sec:timeredef} we define two time variables that allow us to simplify the form of the dynamical equations and examine whether they can be made equal to each other in general. In Sec. \ref{sec:consistency} we analyze the consistency of imposing this equality on the newly defined time variables at least in the asymptotic limit of large black hole masses, and study their relation for finite values of the mass. Finally, we conclude in Sec. \ref{sec:Conclusion} with a discussion of our results. Throughout this article, we set the speed of light and the reduced Planck constant equal to one.

\section{Dynamical equations}\label{sec:eq}

In this section, we investigate an alternative avenue in the computation of the equations that govern the modified dynamics of the interior region of a black hole, based on a more general choice of polymerization parameters off shell. With a suitable choice of lapse function of the form \cite{AOS1,AOS2,AO}
\begin{align}
N=\dfrac{\gamma \delta_b\sqrt{|p_c|}}{\sin\delta_bb},
\end{align}
where $\gamma$ is the Immirzi parameter, the so-called effective Hamiltonian of the system can be written as \cite{AOS1,AOS2,AO}
\begin{align}
NH_{\textrm{eff}}&=\dfrac{L_o}{G}(O_b-O_c),\\
O_b&=-\dfrac{1}{2\gamma}\left(\dfrac{\sin \delta_bb}{\delta_b}+\dfrac{\gamma^2\delta_b}{\sin\delta_bb}\right)\dfrac{p_b}{L_o},\label{Ob}\\
O_c&=\dfrac{1}{\gamma}\dfrac{\sin\delta_cc}{\delta_c}\dfrac{p_c}{L_o}.\label{Oc}
\end{align}
where $L_o$ is the length of the edge parallel to the $x$-direction of the fiducial cell and $G$ is the Newtonian gravitational constant. The canonical variables $b$, $c$, $p_b$, and $p_c$ have the following nonvanishing Poisson brackets:
\begin{align}
\{b,p_b\}=G\gamma,\quad \{c,p_c\}=2G\gamma.
\end{align}
Furthermore, $\delta_b$ and $\delta_c$ are the parameters that capture and introduce the QG effects in the system. The classical Hamiltonian of the model within General Relativity is recovered in the limit $\delta_b,\delta_c\to 0$. It is interesting to note that, should $\delta_b$ and $\delta_c$ be regarded as constants, $N H_{\textrm{eff}}$ is given by the difference of two objects that generate the dynamics in two distinct subsectors of the phase space that are dynamically decoupled. For this reason, we often refer to $O_b$ and $O_c$ as the partial Hamiltonians
\footnote{Strictly speaking, the objects that generate the dynamics in each subsector are $L_oO_b/G$ and $L_oO_c/G$. However, in practice, we will ignore the constant factor $L_o/G$ (which could be reabsorbed through an appropriate redefinition of the lapse function), focus on the more interesting phase space dependent parts, $O_b$ and $O_c$, and use this terminology to refer to them succinctly.}
 of the $b$ and $c$ subsectors, respectively. Both $O_b$ and $O_c$ turn out to be Dirac observables, i.e., constant along each dynamical trajectory. The vanishing of the Hamiltonian constraint implies that, on shell, they are equal to the same constant of motion, $m$, which happens to be proportional to the ADM mass of the black hole.

In the AOS black hole model \cite{AOS1,AOS2,AO}, the parameters $\delta_b$ and $\delta_c$ are treated as constants on the whole phase space and then fixed to have the same value as certain functions of the Dirac observable $m$. The authors of Ref. \cite{N} propose an alternative way to treat these parameters as constants of motion from the very beginning: they take $\delta_b$ and $\delta_c$ as functions of their respective partial Hamiltonian, $\delta_b=\delta_b(O_b)$ and $\delta_c=\delta_c(O_c)$, so that the parameters remain functions of $m$ on shell. Nonetheless, since both $O_b$ and $O_c$ have the same on-shell value, it can be argued that each of the considered parameters should be assumed to be a function of both partial Hamiltonians: on shell, the contribution of one cannot be told apart from that of the other. In this work, we will follow this line of reasoning and investigate its consequences within the ensuing Hamiltonian calculation.

Thus, let $\delta_b$ and $\delta_c$ be functions of both partial Hamiltonians
\begin{align}
\delta_i=f_i(O_b,O_c),
\end{align}
with $i=b,c$. This cross-dependence introduces a coupling between the $b$ and $c$ subsectors that was absent in previous works and that will obviously influence the form of the dynamical equations. Let us begin by computing the equations of motion associated with the connection variables $b$ and $c$, that we will collectively denote with the symbol $i$ (no confusion with the imaginary number will arise in our calculations). We have
\begin{align}
\partial_t i=s_i\dfrac{L_o}{G}\left(\{i,O_i\}-\{i,O_j\}\right),\label{dti}
\end{align}
where $t$ is the time variable associated with the choice of lapse $N$, $i,j=b,c$, $j\neq i$, and $s_i$ is a sign defined by
\begin{align}
s_b=+1,\quad s_c=-1.
\end{align}
The Poisson bracket of $i$ with its respective partial Hamiltonian $O_i$ is given by
\begin{align}
\{i,O_i\}=\{i,p_i\}\dfrac{\partial O_i}{\partial p_i}+\dfrac{\partial O_i}{\partial \delta_i}\left(\dfrac{\partial f_i}{\partial O_i}\{i,O_i\}+\dfrac{\partial f_i}{\partial O_j}\{i,O_j\}\right).\label{eq1}
\end{align}
Similarly,
\begin{align}
\{i,O_j\}=\dfrac{\partial O_j}{\partial \delta_j}\left(\dfrac{\partial f_j}{\partial O_i}\{i,O_i\}+\dfrac{\partial f_j}{\partial O_j}\{i,O_j\}\right).\label{eq2}
\end{align}
The Poisson brackets of the connection variables with each partial Hamiltonian can be solved for in the system of linear equations formed by Eqs. \eqref{eq1} and \eqref{eq2}. When rewritten appropriately, this system can be recast in matrix form: 
\begin{align}
\left(\begin{matrix}
1-\Delta_{ii}&-\Delta_{ij}\\
-\Delta_{ji}&1-\Delta_{jj}
\end{matrix}
\right)\left(\begin{matrix}\{i,O_i\}\\\{i,O_j\}\end{matrix}\right)=\left(\begin{matrix}\{i,p_i\}\dfrac{\partial O_i}{\partial p_i}\\0\end{matrix}\right),\label{sistmat}
\end{align}
where we have defined
\begin{align}
\Delta_{ij}=\dfrac{\partial O_i}{\partial \delta_i}\dfrac{\partial f_i}{\partial O_j}.\label{deltaij}
\end{align}
The system \eqref{sistmat} can be solved if and only if 
\begin{align}
(1-\Delta_{ii})(1-\Delta_{jj})-\Delta_{ij}\Delta_{ji}\neq 0.\label{invertibility}
\end{align}
Assuming that this invertibility condition holds,
\begin{align}
\left(\begin{matrix}\{i,O_i\}\\\{i,O_j\}\end{matrix}\right)=\dfrac{\{i,p_i\}\dfrac{\partial O_i}{\partial p_i}}{(1-\Delta_{ii})(1-\Delta_{jj})-\Delta_{ij}\Delta_{ji}}\left(\begin{matrix}1-\Delta_{jj}\\\Delta_{ji}\end{matrix}\right).
\end{align}
Therefore, by virtue of Eq. \eqref{dti},
\begin{align}
\partial_t i=\dfrac{1-\Delta_{jj}-\Delta_{ji}}{(1-\Delta_{ii})(1-\Delta_{jj})-\Delta_{ij}\Delta_{ji}}\left[s_i\dfrac{L_o}{G}\{i,p_i\}\dfrac{\partial O_i}{\partial p_i}\right],\label{dtifinal}
\end{align}
with $i,j=b,c$ and $j\neq i$. Following the same reasoning, the equations of motion associated with the triad variables $p_b$ and $p_c$ turn out to be
\begin{align}
\partial_t p_i=\dfrac{1-\Delta_{jj}-\Delta_{ji}}{(1-\Delta_{ii})(1-\Delta_{jj})-\Delta_{ij}\Delta_{ji}}\left[-s_i\dfrac{L_o}{G}\{i,p_i\}\dfrac{\partial O_i}{\partial i}\right],\label{dtpifinal}
\end{align}
with $i,j=b,c$ and $j\neq i$.

It is worth noting that the objects in square brackets in Eqs. \eqref{dtifinal} and \eqref{dtpifinal} are the dynamical equations that result when $\delta_b$ and $\delta_c$ are treated as constants on the whole phase space, i.e. those used in Refs. \cite{AOS1,AOS2,AO}. Therefore, if we allow the quantum parameters to be functions of both partial Hamiltonians, the equations of motion are modified via a multiplicative phase space dependent factor,
\begin{align}
C_{ij}=\dfrac{1-\Delta_{jj}-\Delta_{ji}}{(1-\Delta_{ii})(1-\Delta_{jj})-\Delta_{ij}\Delta_{ji}}.
\end{align}
As expected, this factor reduces to the one found in Ref. \cite{N} when the $b$ and $c$ subsectors are decoupled. Indeed, in that case $\Delta_{ij}=0$ if $i\neq j$, and $C_{ij}$ reduces to 
\begin{align}
C_{ij}\to C_i=\dfrac{1}{1-\Delta_{ii}},
\end{align}
which is identical to what the authors of that reference called $F_i^{-1}$.

\section{Time redefinitions}\label{sec:timeredef}

Let $\vec{v}_{H_0}$ be the Hamiltonian vector field associated with the Hamiltonian $H_0$ that is identical to $NH_{\textrm{eff}}$ except for the fact that $\delta_b$ and $\delta_c$ are constant on the whole phase space,
\begin{align}
\vec{v}_{H_0}=\left(\dfrac{\partial H_0}{\partial p_b},-\dfrac{\partial H_0}{\partial b},\dfrac{\partial H_0}{\partial p_c},-\dfrac{\partial H_0}{\partial c}\right)=\left(\vec{v}_{H_{0,b}},\vec{v}_{H_{0,c}}\right).
\end{align}
According to the results of the previous section, when the parameters of the model are instead given by functions of both $O_b$ and $O_c$, the Hamiltonian vector field is given by
\begin{align}
\vec{v}_{H_{\textrm{eff}}}=\left(C_{bc}\vec{v}_{H_{0,b}},C_{cb}\vec{v}_{H_{0,c}}\right).
\end{align}
This local rescaling of the Hamiltonian vector field implies that it is possible to introduce a suitable redefinition of the time variable in each subsector such that one can recover the simpler dynamics generated by $H_0$. However, the fact that $C_{ij}$ is, in general, nonsymmetric means that this change of time is different in the $b$ and $c$ subsectors of the phase space. Indeed, it is immediate to see that, with the appropriate time redefinitions, the dynamical equations become 
\begin{align}
\partial_{\tilde{t}_i}i=s_i\dfrac{L_o}{G}\{i,p_i\}\dfrac{\partial O_i}{\partial p_i},\quad \partial_{\tilde{t}_i}p_i=-s_i\dfrac{L_o}{G}\{i,p_i\}\dfrac{\partial O_i}{\partial i},\label{simpledyneq}
\end{align}
where the sector-dependent time variable $\tilde{t}_i$ is defined in the following manner:
\begin{align}
d\tilde{t}_i=C_{ij}dt,\label{timeredef}
\end{align}
with $i,j=b,c$ and $j\neq i$. Hence, we see that the dynamics that we obtain in the $b$ and $c$ subsectors coincides with that of the AOS model \cite{AOS1,AOS2,AO} when the corresponding equations of motion \eqref{dtifinal} and \eqref{dtpifinal} are rewritten in terms of two time variables, $\tilde{t}_b$ and $\tilde{t}_c$, which are in general \emph{different in each subsector}. This observation provides a strategy to solve the dynamical equations obtained in Sec. \ref{sec:eq}: perform the time redefinitions $t\to\tilde{t}_i$, solve the resulting simpler equations of motion and rewrite the solutions in terms of the original time variable through the integration of Eq. \eqref{timeredef}.

An appealing possibility that we are going to study is whether these time variables can be set to be equal by making use of the freedom that exists off shell. Let us assume that, on shell, $\tilde{t}_b=\alpha \tilde{t}_c$, where $\alpha$ is a real constant. This directly implies that
\begin{align}
C_{bc}|_{\textrm{on}}=\alpha C_{cb}|_{\textrm{on}} \Rightarrow 1-\Delta_{cc}|_{\textrm{on}}-\Delta_{cb}|_{\textrm{on}}=\alpha(1-\Delta_{bb}|_{\textrm{on}}-\Delta_{bc}|_{\textrm{on}}), \label{1cond}
\end{align}
where the symbol $|_{\textrm{on}}$ denotes on-shell evaluation, i.e. evaluation on the phase space region where $H_{\textrm{eff}}=0$. This requirement constitutes a restriction in the form of the first derivatives of the polymerization parameters with respect to the partial Hamiltonians. In the case $\alpha=1$ (of direct application to the AOS model), this condition reduces to
\begin{align}
\Delta_{bb}|_{\textrm{on}}+\Delta_{bc}|_{\textrm{on}}=\Delta_{cc}|_{\textrm{on}}+\Delta_{cb}|_{\textrm{on}}.
\end{align}
We will however consider an arbitrary value of $\alpha$. Rewriting Eq. \eqref{1cond} by using the definition of $\Delta_{ij}$ \eqref{deltaij}, we obtain that the following condition must be satisfied:
\begin{align}
1-\dfrac{\partial O_c}{\partial \delta_c}\bigg|_{\textrm{on}}\left(\dfrac{\partial f_c}{\partial O_c}\bigg|_{\textrm{on}}+\dfrac{\partial f_c}{\partial O_b}\bigg|_{\textrm{on}}\right)=\alpha\left[1-\dfrac{\partial O_b}{\partial \delta_b}\bigg|_{\textrm{on}}\left(\dfrac{\partial f_b}{\partial O_b}\bigg|_{\textrm{on}}+\dfrac{\partial f_b}{\partial O_c}\bigg|_{\textrm{on}}\right)\right].\label{2cond}
\end{align}
Since the two parameters are functions only of the partial Hamiltonians, their evaluation on shell amounts to setting $O_b=O_c=m$ in $f_i$. However, it will prove more useful to rewrite $\delta_i$ as functions of the linear combinations
\begin{align}
\mu_1=\dfrac{O_b+O_c}{2},\quad \mu_2=\dfrac{O_b-O_c}{2},
\end{align}
the on-shell values of which are given by $\mu_1|_{\textrm{on}}=m$ and $\mu_2|_{\textrm{on}}=0$. Then, Eq. \eqref{2cond} can be rewritten as
\begin{align}
1-\dfrac{\partial O_c}{\partial \delta_c}\bigg|_{\textrm{on}}\dfrac{\partial f_c}{\partial \mu_1}\bigg|_{\textrm{on}}=\alpha\left(1-\dfrac{\partial O_b}{\partial \delta_b}\bigg|_{\textrm{on}}\dfrac{\partial f_b}{\partial \mu_1}\bigg|_{\textrm{on}}\right).
\end{align}
Assuming that the functions $f_i$ are $\mathcal{C}^1$, we can evaluate them on shell first and then compute the derivatives. Thus, for $\tilde{t}_b$ and $\tilde{t}_c$ to be proportional, it must be satisfied on shell that
\begin{align}
1-\dfrac{\partial O_c}{\partial \delta_c}\bigg|_{\textrm{on}}\dfrac{\partial f_c(m,0)}{\partial m}=\alpha\left(1-\dfrac{\partial O_b}{\partial \delta_b}\bigg|_{\textrm{on}}\dfrac{\partial f_b(m,0)}{\partial m}\right).\label{3cond}
\end{align}
In the rest of our discussion, we will omit the on-shell evaluations in formulas of this kind to simplify our notation. The on-shell restriction will be clear from the context.

The derivatives of the partial Hamiltonians with respect to $\delta_b$ and $\delta_c$ are 
\begin{align}
\dfrac{\partial O_b}{\partial \delta_b}&=-\dfrac{1}{2\gamma}\left(1-\dfrac{\gamma^2\delta_b^2}{\sin^2\delta_bb}\right)\dfrac{\delta_bb\cos\delta_bb-\sin\delta_bb}{\delta_b^2}\dfrac{p_b}{L_o},\label{dObddb}\\
\dfrac{\partial O_c}{\partial \delta_c}&=\dfrac{1}{\gamma}\dfrac{\delta_cc\cos\delta_cc-\sin\delta_cc}{\delta_c^2}\dfrac{p_c}{L_o},\label{dOcddc}
\end{align}
as can be immediately derived from Eqs. \eqref{Ob} and \eqref{Oc}. The fact that these derivatives depend on the canonical variables seems in tension with the requirement that Eq. \eqref{3cond} must be satisfied on the whole phase space. Since the derivatives $\partial_mf_i(m,0)$ only depend on $m$ (and, therefore, remain constant along each dynamical trajectory), the functions $f_i$ need to be selected so that any phase space dependent contribution is canceled identically. 

In order to evaluate these derivatives on shell, it is necessary to identify the independent functional dependences. It is immediate to see that the dependence on the connection variables $b$ and $c$ can be removed in terms of their momenta and the black hole mass. Indeed, the functions of each connection variable can be rewritten in terms of its corresponding partial Hamiltonian and triad variable. Using Eqs. \eqref{Ob} and \eqref{Oc},
\begin{align}
\dfrac{\sin\delta_bb}{\delta_b}&=-\dfrac{\gamma L_o O_b}{p_b}\left(1+\sqrt{1-\dfrac{p_b^2}{L_o^2O_b^2}}\right),\\
\dfrac{\sin\delta_cc}{\delta_c}&=\dfrac{\gamma L_o O_c}{p_c}.
\end{align}
Hence, the only independent functional dependences that remain on shell are those associated with the triad variables. By means of the above relations, we can recast every function of $b$ and $c$ that appears in Eqs. \eqref{dObddb} and \eqref{dOcddc} as a function of $p_b$, $p_c$, and the partial Hamiltonians, which reduce to $m$ after the on-shell evaluation. After a straightforward computation, we obtain on shell that 
\begin{align}
\dfrac{\partial O_c}{\partial \delta_c}=\dfrac{\pm \arcsin\left[\dfrac{\gamma L_o mf_c(m,0)}{p_c}\right]\sqrt{\dfrac{p_c^2}{L_o^2}-\gamma^2m^2f_c^2(m,0)}-\gamma m f_c(m,0)}{\gamma f_c^2(m,0)},
\end{align}
where the sign $\pm$ corresponds with the sign of $\cos \delta_cc$. A similar, although more complicated expression can be found for the on-shell value of $\partial O_b/\partial \delta_b$. To shorten this expression, we use (exclusively here) the compact notation $\tilde{p}_b= p_b/(L_o m)$:

\begin{align}
\dfrac{\partial O_b}{\partial \delta_b}&=\dfrac{m}{2\gamma f_b^2(m,0)}\left[1-{\tilde{p}_b^2}\left(1+\sqrt{1-{\tilde{p}_b^2}}\right)^{-2}\right]\left\{-\gamma f_b(m,0)\left(1+\sqrt{1-{\tilde{p}_b^2}}\right)\right.\nonumber\\
&\left.\pm\arcsin\left[\dfrac{\gamma f_b(m,0)}{\tilde{p}_b}\left(1+\sqrt{1-{\tilde{p}_b^2}}\right)\right]\sqrt{{\tilde{p}_b^2}-\gamma^2f_b^2(m,0)\left(1+\sqrt{1-{\tilde{p}_b^2}}\right)^2}\right\}. 
\end{align}

On the light of these relations, we realize that the condition \eqref{3cond} has the following structure:
\begin{align}
1-F_c(p_c)\dfrac{\partial f_c(m,0)}{\partial m}=\alpha\left[1-F_b(p_b)\dfrac{\partial f_b(m,0)}{\partial m}\right],
\end{align}
where the functional forms of $F_b$ and $F_c$ are irrelevant for the present argument, except that they are not constant functions. For this condition to hold on the whole phase space, the derivatives of the polymerization parameters must vanish
\begin{align}
\dfrac{\partial f_i(m,0)}{\partial m}=0,
\end{align}
and $\alpha$ must be equal to one. As a result, we realize that, if we demand that $\tilde{t}_b$ and $\tilde{t_c}$ be proportional for all values of the mass, the only possibility is that they are equal and the polymerization parameters are constants. In other words, we cannot reconcile the choice of these parameters as Dirac observables and the dynamics being governed by Eqs. \eqref{simpledyneq} in a single time variable, at least for all values of the mass. The conclusion to be drawn from this result is that the appearance of two distinct time variables that simplify the dynamics in the radial and angular subsectors of phase space is a defining feature of the model so long as treating the parameters $\delta_i$ as proper Dirac observables is required. In the next section, we will examine whether this condition can be imposed consistently for black hole masses much larger than the Planck mass. 

\section{Consistency in the limit of large black hole masses}\label{sec:consistency}

In this section, we investigate whether the dynamics of the AOS model, that results from considering the parameters $\delta_i$ as constant numbers in the Hamiltonian calculations, can be recovered at least in the limit of large black hole masses when these parameters are taken instead as constants of motion and one introduces a convenient time redefinition. Let us begin by assuming that it is possible. Then, in the considered limit, the dynamical equations adopt the same form as in Refs. \cite{AOS1,AOS2,AO}, up to subdominant terms that reflect the fact that $\tilde{t}_b$ and $\tilde{t}_c$ cannot be made equal for all values of the mass. Furthermore, according to the argument presented in Refs. \cite{AOS1,AOS2,AO}, the parameters $\delta_i$ are found to be given by
\begin{align}
\delta_b=\left(\dfrac{\sqrt{\Delta}}{\sqrt{2\pi}\gamma^2m}\right)^{1/3}+o(m^{-1/3}),\quad \delta_c=\dfrac{1}{2L_o}\left(\dfrac{\gamma \Delta^2}{4\pi^2m}\right)^{1/3}+o(m^{-1/3}),\label{deltas}
\end{align}
where the symbol $o(\cdot)$ collectively denotes all the terms that are subdominant with respect to the function inside the parentheses. In these expressions, $\Delta$ is the area gap in loop quantum gravity. Therefore, asymptotically,
\begin{align}
\dfrac{\partial f_i}{\partial m}=-\dfrac{1}{3}\dfrac{\delta_i}{m}+o\left(\dfrac{\delta_i}{m}\right),\label{derivativedeltas}
\end{align}
with $i=b,c$. Then, the condition \eqref{3cond} can be imposed consistently as long as\footnote{Should any of the exponents $n_i$ be equal to $4/3$, it is straightforward to realize that a further condition on the coefficients of the dominant terms must be met for Eq. \eqref{3cond} to hold asymptotically.}

\begin{align}
\lim_{m\to\infty}\dfrac{\partial O_i}{\partial \delta_i}\sim m^{n_i},\quad n_i\leq \dfrac{4}{3},
\end{align}
with $i=b,c$.

Given that we are working under the assumption that $\tilde{t}_b=\tilde{t_c}=\tilde{t}$ when $m\to\infty$, we obtain from the solutions of Refs. \cite{AOS1,AOS2,AO} that, in this limit and up to subdominant corrections,
\begin{align}
\cos\delta_bb(\tilde{t})&=b_o\tanh\left[\dfrac{1}{2}\left(b_o\tilde{t}+2\tanh^{-1}\dfrac{1}{b_o}\right)\right],\label{beff}\\
\tan\dfrac{\delta_c c(\tilde{t})}{2}&=\dfrac{\gamma L_o \delta_c}{8m}e^{-2\tilde{t}},\label{ceff}\\
p_b(\tilde{t})&=-2\dfrac{\sin\delta_cc(\tilde{t})}{\delta_c}\dfrac{\sin\delta_bb(\tilde{t})}{\delta_b}\dfrac{p_c(\tilde{t})}{\gamma^2+\dfrac{\sin^2\delta_bb(\tilde{t})}{\delta_b^2}},\label{pbeff}\\
p_c(\tilde{t})&=4m^2\left(e^{2\tilde{t}}+\dfrac{\gamma^2 L_o^2\delta_c^2}{64m^2}e^{-2\tilde{t}}\right),\label{pceff}
\end{align}
where $b_o=\sqrt{1+\gamma^2\delta_b^2}$. Let us now proceed to the computation of the dominant terms of $\partial O_i/\partial \delta_i$.

From Eq. \eqref{pceff}, it follows immediately that
\begin{align}
p_c=4e^{2\tilde{t}}m^2+o\left(m^2\right).
\end{align}
The case of the connection variable $c$ and its trigonometric functions is less immediate. Since the solution written above involves the tangent of $\delta_cc/2$, it is useful to employ the identity
\begin{align}
\cos \delta_cc=\dfrac{1-\tan^2(\delta_cc/2)}{1+\tan^2(\delta_cc/2)}.
\end{align}
Then, by virtue of Eq. \eqref{ceff},
\begin{align}
\cos\delta_cc=1-\dfrac{\gamma^2 L_o^2}{32e^{4\tilde{t}}}\dfrac{\delta_c^2}{m^2}+o\left(\dfrac{\delta_c^2}{m^2}\right).
\end{align}
Given that the sum of the squares of the sine and cosine functions is equal to one, we can obtain $\sin\delta_cc$ and $\delta_cc$ from the expression above. In order to do this, it suffices to bear in mind that, according to the conventions of Refs. \cite{AOS1,AOS2,AO}, $b>0$, $c>0$, $p_b\leq 0$, and $p_c\geq 0$. This, together with the fact that $\lim_{m\to \infty}\delta_i=0$, implies that every relevant trigonometric function is nonnegative.

After a straightforward calculation, we conclude that
\begin{align}
\dfrac{\partial O_c}{\partial \delta_c}=-\dfrac{\gamma^2L_o^2}{48 e^{4\tilde{t}}}\dfrac{\delta_c}{m}+o\left(\dfrac{\delta_c}{m}\right).\label{dOcddeltacasin}
\end{align}
The dominant term goes with $m^{-4/3}$ [see Eq. \eqref{deltas}], which implies that the left hand side of Eq. \eqref{3cond} tends to one in the limit of large black hole masses. Let us now perform the analogous analysis on the right hand side of Eq. \eqref{3cond}. 

On the light of the form of the solution \eqref{beff} for the connection variable $b$, it proves useful to employ the identity
\begin{align}
\tanh(a+b)=\dfrac{\tanh a+\tanh b}{1+\tanh a \tanh b},
\end{align}
such that, up to subdominant corrections to the leading time-dependent contribution,
\begin{align}
\cos\delta_bb=\dfrac{1+b_o\tanh(b_oT)}{1+b_o^{-1}\tanh(b_oT)},
\end{align}
where we have defined $T=\tilde{t}/2$. Rewriting the previous expression as a power series, we get
\begin{align}
\cos\delta_bb = 1+C_1\delta_b^2+o\left(\delta_b^2\right),
\end{align}
with a constant $C_1$ given by
\begin{align}
C_1&=\gamma^2\dfrac{\tanh T}{1+\tanh T}.
\end{align}
Recasting every function of $b$ that appears in Eq. \eqref{dObddb} as a power series, we find that 
\begin{align}
\left(1-\dfrac{\gamma^2\delta_b^2}{\sin^2\delta_bb}\right)\dfrac{\delta_bb\cos\delta_bb-\sin\delta_bb}{\delta_b^2}=-\dfrac{2}{3}\left(2C_1+\gamma^2\right)\dfrac{C_1}{\sqrt{-2C_1}}\delta_b+o\left(\delta_b\right).
\end{align}
Lastly, the asymptotic value of $p_b$ can be obtained from Eq. \eqref{pbeff} by introducing the appropriate expansions:
\begin{align}
-\dfrac{1}{2}\dfrac{p_b}{\gamma L_o}=\dfrac{\sqrt{-2C_1}}{\gamma^2-2C_1}m+o\left(m\right).
\end{align}
In conclusion, the partial derivative of $O_b$ with respect to $\delta_b$ displays the asymptotic behavior
\begin{align}
\dfrac{\partial O_b}{\partial \delta_b}=-\dfrac{2}{3}C_1\dfrac{\gamma^2+2C_1}{\gamma^2-2C_1}\delta_bm+o\left(\delta_bm\right).
\end{align}
This, in conjunction with Eq. \eqref{derivativedeltas}, leads to the asymptotic vanishing of $\Delta_{bb}+\Delta_{bc}$. Therefore, the condition \eqref{3cond} reduces to $\alpha=1$ in the asymptotic limit $m\rightarrow\infty$. For this value of $\alpha$, the times $\tilde{t}_b$ and $\tilde{t}_c$ indeed coincide for black holes of asymptotically large masses.

From this result, we conclude that, among all possible choices of $\tilde{t}_b$ and $\tilde{t}_c$ such that they are proportional to each other, only the choice where the proportionality constant is equal to one is admissible in the limit of large black hole masses. Therefore, the AOS solutions can at least be reconciled with the present calculation in this limit. It is important to remark that, although the equations of motion do coincide with those derived in Refs. \cite{AOS1,AOS2,AO} in the limit of large black hole masses, they are written in a \emph{different time variable} $\tilde{t}$. Thus, the spacetime geometry is modified with respect to the one studied in those works. This opens a door to a different asymptotic behavior of the spacetime metric of the exterior region, which in particular may have a different asymptotic (flat) behavior \cite{Bouhmadicrit}.  Additionally, the lapse function associated with $\tilde{t}$ will present a different phase space dependence, that may call for a different re-densitization of the Hamiltonian constraint $NH_{\textrm{eff}}$ with respect to the difference $O_b-O_c$ of what we have called the partial Hamiltonians of the model. We leave these issues for future studies and, as we declared in the Introduction, restrict our discussion here to the viability of a Hamiltonian derivation of the AOS solution (possibly in an asymptotic sense) treating the polymerization parameters as true constants of motion.

Let us close this section by studying the relation between the time variables $\tilde{t}_b$ and $\tilde{t}_c$ for finite values of the mass. In view of their definitions \eqref{timeredef} and the fact that the denominator of $C_{ij}$ is symmetric, we obtain that the ratio of the two differential times is 
\begin{align}
\dfrac{d\tilde{t}_b}{d\tilde{t}_c}=\dfrac{1-\Delta_{cc}-\Delta_{cb}}{1-\Delta_{bb}-\Delta_{bc}}.
\end{align}
On shell, this is equivalent to
\begin{align}
\dfrac{d\tilde{t}_b}{d\tilde{t}_c}=\dfrac{1-\dfrac{\partial O_c}{\partial \delta_c}\dfrac{\partial f_c}{\partial m}}{1-\dfrac{\partial O_b}{\partial \delta_b}\dfrac{\partial f_b}{\partial m}}.
\end{align}
Notice that the numerator (denominator) of the right hand side is a function of $\tilde{t}_c$ ($\tilde{t}_b$). Therefore, by integrating, we obtain an equality between a function of $\tilde{t}_b$ and a function of $\tilde{t}_c$, which provides an implicit relation between the two time variables,
\begin{align}
\int_{\tilde{t}_b}^0 d\tilde{t}_b' \left(1-\dfrac{\partial O_b}{\partial \delta_b}\dfrac{\partial f_b}{\partial m}\right)=\int_{\tilde{t}_c}^0 d\tilde{t}_c'\left(1-\dfrac{\partial O_c}{\partial \delta_c}\dfrac{\partial f_c}{\partial m}\right),
\end{align} 
where the choice of integration limits reflects the fact that, according to the conventions of Refs. \cite{AOS1,AOS2,AO}, the time variables are negative in the interior region of the black hole, and their origins coincide. We have already determined that, in the asymptotic limit of large masses, this relation reduces to the identity. However, for finite values of the black hole mass, the difference between both time variables is given by
\begin{align}
\tilde{t}_b-\tilde{t}_c=\dfrac{\partial f_b}{\partial m}\int_{0}^{\tilde{t}_b} \dfrac{\partial O_b}{\partial \delta_b} d\tilde{t}_b' -\dfrac{\partial f_c}{\partial m}\int_{0}^{\tilde{t}_c}  \dfrac{\partial O_c}{\partial \delta_c} d\tilde{t}_c'.
\end{align}
Inserting the results obtained in this section,
\begin{align}
\tilde{t}_b-\tilde{t}_c=&\dfrac{\partial f_b}{\partial m}\int_0^{\tilde{t}_b} d\tilde{t}_b' \left[-\dfrac{2}{3}C_1(\tilde{t}_b')\dfrac{\gamma^2+2C_1(\tilde{t}_b')}{\gamma^2-2C_1(\tilde{t}_b')}\delta_bm+o\left(\delta_bm\right)\right]\nonumber\\
&-\dfrac{\partial f_c}{\partial m} \int_0^{\tilde{t}_c} d\tilde{t}_c' \left[-\dfrac{\gamma^2L_o^2}{48 e^{4\tilde{t}'_c}}\dfrac{\delta_c}{m}+o\left(\dfrac{\delta_c}{m}\right)\right].
\end{align}
Here, we have used that, at the order of approximation needed in the integrals of our expression in the asymptotic limit of large masses, we can identify $\tilde{t}_b=\tilde{t}_c=\tilde{t}$. Since the dominant term of the second integral, which goes as $\delta_c/m$, is already subdominant with respect to that of the first integral, only the $b$ sector contributes to the studied difference at the lowest nontrivial order. We have
\begin{align}
\tilde{t}_b-\tilde{t}_c=-\dfrac{4}{3}\gamma^2\dfrac{\partial f_b}{\partial m}\left[\int_0^{\tilde{t}_b/2}\dfrac{\tanh T_b(1+3\tanh T_b)}{1-\tanh^2 T_b}d T_b\right]\delta_bm+o\left(\delta_b^2\right),
\end{align}
so that
\begin{align}\label{tc(tb)}
\tilde{t}_c=\tilde{t}_b-\dfrac{1}{9}\gamma^2\left(-3\tilde{t}_b+3\sinh \tilde{t}_b+\cosh \tilde{t}_b-1\right)\delta_b^2+o\left(\delta_b^2\right).
\end{align}
This equation provides the first-order corrected relation between both time variables, their difference vanishing when $m\to\infty$, as expected. Additionally, this relation reveals another property that was already pointed out in Ref. \cite{N}: the difference $\tilde{t}_c-\tilde{t}_b$ also vanishes in the region where quantum effects are negligible, i.e., close to the event horizon. When solving the equations of motion associated with the $b$ sector, the constants of integration were fixed in such a way that the horizon (defined by $b=0$ and $p_b=0$) lies at $\tilde{t}_b=0$ \cite{AOS1,AOS2,AO}. It is immediate to verify that both time variables are indeed close to each other when $\tilde{t}_b$ approaches zero, since the dominant term of their difference vanishes at least as $\tilde{t}_b^2$ in this limit, when asymptotically large masses are considered. 

The dominant-order correction to the difference of times in Eq. \eqref{tc(tb)} also makes it apparent that the relation between $\tilde{t}_b$ and $\tilde{t}_c$ may become nonmonotonic in general, which allows us to draw yet another parallel with Ref. \cite{N}. Indeed,
\begin{align}
\dfrac{d\tilde{t}_c}{d\tilde{t}_b}=1-\dfrac{1}{9}\gamma^2(\sinh\tilde{t}_b+3\cosh\tilde{t}_b-3)\delta_b^2+o\left(\delta_b^2\right)
\end{align}
is positive in the limit $\tilde{t}_b\to 0$ but may reach a value of $\tilde{t}_b$ where it vanishes and, eventually, changes sign. For the standard value of the Immirzi parameter $\gamma=0.2375$, $\Delta=4\sqrt{3}\pi G\gamma$, and $m=10000 m_{\textrm{Pl}}$ (where $m_{\textrm{Pl}}$ denotes the Planck mass), we obtain that this derivative vanishes at $\tilde{t}_b\approx -9.36 t_{\textrm{Pl}}$ (where $t_{\textrm{Pl}}$ is the Planck time). Therefore, while  $\tilde{t}_c$ decreases as $\tilde{t}_b$ decreases near the horizon, this trend is found to be reversed beyond a critical value of $\tilde{t}_b$ at the considered truncation order in the asymptotic expansion. This indicates that there would exist a point along the evolution where the ratio $C_{bc}/C_{cb}$ would cease to be finite. Nonetheless, note that this does not necessarily imply that the invertibility condition \eqref{invertibility} would be violated at that point, not even at our truncation order. Actually, the symmetry properties of $C_{ij}$ ensure that only the numerators of $C_{bc}$ and $C_{cb}$ contribute to their ratio. As a result, the nonfiniteness of this ratio at a certain point along the evolution should not be attributed in principle to an ill behavior of the denominator of $C_{ij}$ and, consequently, to a violation of condition \eqref{invertibility}.

\section{Conclusion}\label{sec:Conclusion}

In this paper, we have examined whether it is possible to construct a Hamiltonian formalism where the polymerization parameters that encode the quantum corrections in black hole spacetimes can be treated as constants of motion. The final identification of these parameters with dynamical constants is one of the ideas of the AOS model, proposed in Refs. \cite{AOS1,AOS2,AO}. However, instead of incorporating this identification into the Hamiltonian calculation from the beginning, the analysis in those references is carried out ignoring the Poisson brackets of the parameters, treating them as constants on the whole phase space. It is only later on that their value is set equal to certain functions of the black hole mass, which is a Dirac observable of the system under consideration. The authors of Ref. \cite{N} pointed out that the computation of the Hamiltonian equations would change if one takes into consideration those Poisson brackets, regarding the parameters as true constants of motion. To show this, it was noticed in Ref. \cite{N} that, given the form of the Hamiltonian, there are two dynamically decoupled subsectors in phase space, provided that the polymerization parameters do not introduce any cross-dependence. With this caveat, each subsector can be studied separately and its dynamics is generated by one of the two terms that appear in the Hamiltonian constraint (with a suitable choice of lapse). We have referred to these two terms as \emph{partial Hamiltonians}, which turn out to be Dirac observables that reduce to the black hole mass on shell. Imposing that the polymerization parameter associated with each subsector is a function of its corresponding partial Hamiltonian, the equations of motion that one obtains differ from those of the AOS model by a phase space dependent factor that complicates the solutions. However, this factor can be reabsorbed by appropriate time redefinitions, leading to simpler dynamical equations written in two separate time variables, one in each subsector. In Ref.  \cite{N}, both variables were found to be approximately equal from the event horizon up to a neighborhood of the transition surface where QG effects become important, concluding that the results of the AOS model were approximately valid when restricted to this region of the interior of the black hole. 

In the present work, we have extended the aforementioned analysis to take into account the possibility that the polymerization parameters, regarded as constants of motion, depend not only on their corresponding partial Hamiltonian, but on \emph{both} of them. This possibility breaks the decoupling of subsectors that plays a central role in Ref. \cite{N}. Indeed, since both partial Hamiltonians coincide with the value of the black hole mass on shell by virtue of the vanishing of the Hamiltonian constraint, one should in principle not be able to tell their contributions apart. A dependence on both of these Dirac observables brings new freedom to the treatment of the polymerization parameters. We have investigated whether this new off-shell freedom can help to derive the AOS model exclusively from a standard Hamiltonian calculation, viewing the parameters as functions of both Dirac observables from the beginning. We have derived in Sec. \ref{sec:eq} the corresponding equations of motion that govern the dynamics in the interior region. These equations turn out to be corrected by a phase space dependent factor as well, although its functional form is complicated by the fact that the two subsectors no longer decouple dynamically. We have observed that this factor does reduce to the one found in Ref. \cite{N} in the limit where the decoupling is recovered. In Sec. \ref{sec:timeredef}, we have written down the time redefinitions that allow us to simplify the dynamics, leading to equations of motion that are identical to those that result from considering constant parameters, although now written in two different time variables. We have then discussed whether these newly defined time variables can be required to be equal to each other. Remarkably, the answer turns out to be in the negative in spite of the commented off-shell freedom, since this condition would imply that the polymerization parameters are necessarily constants on the whole phase space. In Sec. \ref{sec:consistency}, we have verified whether this equality of time variables can be imposed at least in the limit of infinitely large black hole masses, as one would expect to be the case in order to recover the standard results of General Relativity in this asymptotic limit. Indeed, we have proven that one can require this coincidence of times consistently. We have also studied the first-order correction to the relation between both time variables, which has allowed us to draw parallels with previous results obtained in Ref. \cite{N}. First, the two time variables are still approximately similar to each other near the event horizon, where the QG effects are not relevant. Second, for finite rather than asymptotically large black hole masses, the dynamical solutions are such that a point in the evolution may generically be reached where the time flow would be reversed, in the sense that the relation between the two time variables would not be monotonic around it.

Our conclusions imply that the results of Refs. \cite{AOS1,AOS2,AO}, which are based on Hamiltonian calculations where the polymerization parameters are treated as constant numbers, can be \emph{partially} reconciled with a treatment where these parameters are regarded as proper constants of motion, at least for black holes with large masses, which on the other hand are the focus of the analysis of those references. The wording ``partially'' is key here. In particular, one should not forget that the spacetime geometry is modified with respect to that of the AOS model by means of time redefinitions. Even if this apparently slight modification does not alter some of the conclusions of  Refs. \cite{AOS1,AOS2,AO}, it may affect e.g. the rate at which the metric decays at spatial infinity.\footnote{Although we have not explicitly dealt with the exterior region in this article, the same procedure is applicable to that case.} This matter will constitute the subject of future work. 

\acknowledgments
	
The authors are very grateful to B. Elizaga Navascu\'es for discussions. This work has been supported by Project. No. MICINN FIS2017-86497-C2-2-P from Spain (with extension Project. No. MICINN PID2020-118159GB-C41 under evaluation). The project that gave rise to these results received the support of a fellowship from ``la Caixa'' Foundation (ID 100010434). The fellowship code is LCF/BQ/DR19/11740028. Partial funds for open access publication have been received from CSIC.


\begin{thebibliography}{100}

\bibitem{AOS1} A. Ashtekar, J. Olmedo, and P. Singh, Quantum transfiguration of Kruskal black holes, Phys. Rev. Lett. \textbf{121}, 241301 (2018).
\bibitem{AOS2} A. Ashtekar, J. Olmedo, and P. Singh, Quantum extension of the Kruskal spacetime, Phys. Rev. D \textbf{98}, 126003 (2018).
\bibitem{AO} A. Ashtekar and J. Olmedo, Properties of a recent quantum extension of the Kruskal geometry, Int. J. Mod. Phys. D \textbf{29}, 2050076 (2020).

\bibitem{1} A. Ashtekar and M. Bojowald, Black hole evaporation: A paradigm, Class. Quantum Grav. \textbf{22}, 3349(2005).
\bibitem{2} A. Ashtekar and M. Bojowald, Quantum geometry and the Schwarzschild singularity, Class. Quantum Grav. \textbf{23}, 391 (2005).
\bibitem{3} D. Cartin and G. Khanna, Wave functions for the Schwarzschild black hole interior, Phys. Rev. D \textbf{73} 104009 (2006).
\bibitem{4} L. Modesto, Loop quantum black hole, Class. Quantum Grav. \textbf{23}, 5587 (2006).
\bibitem{5} M. Bojowald, D. Cartin, and G. Khanna, Lattice refining loop quantum cosmology, anisotropic models and stability, Phys. Rev. D \textbf{76}, 064018 (2007).
\bibitem{6} C. G. Boehmer and K. Vandersloot, Loop quantum dynamics of Schwarzschild interior, Phys. Rev. D \textbf{76}, 1004030 (2007).
\bibitem{7} M. Campiglia, R. Gambini, and J. Pullin, Loop quantization of a spherically symmetric midsuperspaces: The interior problem, AIP Conf. Proc. \textbf{977}, 52 (2008).
\bibitem{8} S. Sabharwal and G. Khanna, Numerical solutions to lattice-refined models in loop quantum cosmology, Class. Quantum Grav. \textbf{25}, 085009 (2008).
\bibitem{9} D. W. Chiou, Phenomenological loop quantum geometry of the Schwarzschild black hole, Phys. Rev. D \textbf{78}, 064040 (2008).
\bibitem{10} D. W. Chiou, Phenomenological dynamics of loop quantum cosmology in Kantowski-Sachs spacetime, Phys. Rev. D \textbf{78}, 044019 (2008).
\bibitem{11} J. Brannlund, S. Kloster, and A. DeBenedictis, The evolution of black holes in the mini-superspace approximation of loop quantum gravity, Phys. Rev. D \textbf{79}, 084023 (2009).
\bibitem{12} R. Gambini, J. Omedo, and J. Pullin, Quantum black holes in loop quantum gravity, Class. Quantum Grav. \textbf{31}, 095009 (2014).
\bibitem{13} R. Gambini and J. Pullin, Hawking radiation from a spherical loop quantum gravity black hole, Class. Quantum Grav. \textbf{31}, 115003 (2014).
\bibitem{14} N. Dadhich, A. Joe, and P. Singh, Emergence of the product of constant curvature spaces in loop quantum cosmology, Class. Quantum Grav. \textbf{32}, 185006 (2015).
\bibitem{15} H. M. Haggard and C. Rovelli, Quantum-gravity effects outside the horizon spark black to white hole tunneling, Phys. Rev. D \textbf{92}, 104020 (2015).
\bibitem{16} A. Joe and P. Singh, Kantowski-Sachs spacetime in loop quantum cosmology: Bounds on expansion and shear scalars and viability of quantization prescriptions, Class. Quantum Grav. \textbf{32}, 015009 (2015).
\bibitem{17} A. Corichi and P. Singh, Loop quantum dynamics of Schwarzschild interior revisited, Class. Quantum Grav. \textbf{33}, 055006 (2016).
\bibitem{18} M. Campiglia, R. Gambini, J. Olmedo, and J. Pullin, Quantum self-gravitating collapsing matter in a quantum geometry, Class. Quantum Grav. \textbf{33}, 18LT01 (2016).
\bibitem{19} S. Saini and P. Singh, Geodesic completeness and the lack of strong singularities in effective loop quantum Kantowski-Sachs spacetime, Class. Quantum Grav. \textbf{33}, 245019 (2016).
\bibitem{20} J. Cortez, W. Cuervo, H. A. Morales-T\'ecotl, and J. C. Ruelas, On effective loop quantum geometry of Schwarzschild interior, Phys. Rev. D \textbf{95}, 064041 (2017).
\bibitem{21} J. Olmedo, S. Saini, and P. Singh, From black holes to white holes: A quantum gravitational symmetric bounce, Class. Quantum Grav. \textbf{34}, 225011 (2017).
\bibitem{22} A. Yonika, G. Khanna, and P. Singh, Von-Neumann stability and singularity resolution in loop quantized Schwarzschild black hole, Class. Quantum Grav. \textbf{35}, 045007 (2018).
\bibitem{23} E. Bianchi, M. Christodoulou, F. D'Ambrosio, H. M. Haggard, and C. Rovelli, White holes as remnants: A surprising scenario for the end of a black hole, Class. Quantum Grav. \textbf{35}, 225003 (2018).
\bibitem{24} N. Bodendorfer, F. M. Mele, and J. M\"unch, Effective quantum extended spacetime of polymer Schwarzschild black hole, Class. Quantum Grav. \textbf{36}, 195015 (2019).
\bibitem{25} E. Alesci, S. Bahrami, and D. Pranzetti, Quantum gravity predictions for black hole interior geometry, Phys. Lett. B \textbf{797}, 134908 (2019).
\bibitem{26} M. Bouhmadi-L\'opez, S. Brahma, C.-Y. Chen, P. Chen, and D.-h- Yeom, A consistent model of non-singular Schwarzschild black hole in loop quantum gravity and its quasinormal modes, J. Cosmol. Astropart. Phys. \textbf{07}, 066 (2020).
\bibitem{27} M. Bojowald, Black-hole models in loop quantum gravity, Universe \textbf{6}, 125 (2020).
\bibitem{28} J. Ben Achour, S. Brahma, S. Mukohyama, and J.-P. Uzan, Towards consistent black-to-white hole bounces from matter collapse, J. Cosmol. Astropart. Phys. \textbf{09}, 020 (2020).
\bibitem{29} R. Gambini, J. Olmedo, and J. Pullin, Spherically symmetric loop quantum gravity: Analysis of improved dynamics,  Class. Quantum Grav. \textbf{37}, 205012 (2020).
\bibitem{30} J. G. Kelly, R. Santacruz, and E. Wilson-Ewing, Effective loop quantum gravity framework for vacuum spherically symmetric spacetimes, Phys. Rev. D \textbf{102}, 106024 (2020).
\bibitem{31} W.-C. Gan, N. O. Santos, F.-W. Shu, and A. Wang, Towards understanding of loop quantum black holes, Phys. Rev. D \textbf{102}, 124030 (2020).
\bibitem{32} J. G. Kelly, R. Santacruz, and E. Wilson-Ewing, Black hole collapse and bounce in effective loop quantum gravity, Class. Quantum Grav. \textbf{38}, 04LT01 (2021).
\bibitem{33} N. Bodendorfer, F. M. Mele, and J. M\"unch, Mass and horizon Dirac observables in effective models of quantum black-to-white hole transition, Class. Quantum Grav. \textbf{38}, 095002 (2021).
\bibitem{34} N. Bodendorfer, F. M. Mele, and J. M\"unch, $(b,v)$-type variables for black to white hole transitions in effective loop quantum gravity, arXiv:1911.12646.
\bibitem{35} R. G. Daghigh, M. D. Green, and G. Kunstatter, Scalar perturbations and stability of a loop quantum corrected Kuskal black hole, arXiv:2012.13359.
\bibitem{36} J. M\"unch, Causal structure of a recent loop quantum gravity black hole collapse model, arXiv:2103.17112.
\bibitem{N} N. Bodendorfer, F. M. Mele, and J. M\"unch, A note on the Hamiltonian as a polymerisation parameter, Class. Quantum Grav. \textbf{37}, 187001 (2019).
\bibitem{Bouhmadicrit} M. Bouhmadi-L\'opez, S. Brahma, C.-Y. Chen, P. Chen, and D.-h- Yeom, Asymptotic non-flatness of an effective black hole model based on loop quantum gravity, Phys. Dark Univ. \textbf{30}, 100701 (2020).
\bibitem{Bojocrit} M. Bojowald, Comment (2) on ``Quantum transfiguration of Kruskal black holes'', arXiv:1906.04650.
\bibitem{Bojo1} M. Bojowald, No-go result for covariance in models of loop quantum gravity, Phys. Rev. D \textbf{102}, 046006 (2020).
\bibitem{Arruga} D. Arruga, J. Ben Achour, and K. Noui, Deformed general relativity and quantum black holes interior, Universe \textbf{6}, 039 (2020).

\end{thebibliography}
\end{document}